\documentstyle[12pt]{article}
\begin{document}
\centerline{SIBR preprint TH-97-S010. February 25, 1997}
\centerline{PACS 03.65.=96w, 3.65.Bz, 03.65.Fd}

\vspace*{2cm}
\begin{center}
{\Large \bf IVARIANT  FORMULATION  OF  q-DEFORMATIONS  VIA  THE  NOVEL
GENOMATHEMATICS}
\end{center}

\vskip 1cm
\begin{center}
{\it {\bf  Ruggero Maria  Santilli}\\  Institute for  Basic Research\\
P. O. Box  1577, Palm Harbor,  FL 34682,  U.S.A.}\\ {\sl ibr@gte.net,
http://home1.gte.net/ibr}
\end{center}

\vskip 1.5cm
\begin{abstract}
In this note we outline the history of q-deformations; indicate their
physical shortcomings; suggest their apparent resolution via an
invariant formulation based on a new mathematics of genotopic type; and
point out their expected physical significance once formulated in an
invariant form.
\end{abstract}

\section{INTRODUCTION}
\setcounter{equation}{0}
        In 1948 Albert [1] introduced the notions of {\it
Lie-admissible} and
{\it Jordan-admissible algebras}  as generally nonassociative algebras
$U$ with
elements $a, b, c,$ and abstract product ab which are such that the
attached algebras $U^-$ and $U^+$, which are the same vector spaces as
$U$
equipped with the products  $[a, b]_U \;=\; ab - ba$ and $ \{a, b\}_U
\;=\;
ab + ba$, are Lie and Jordan algebras, respectively.  Albert then
studied the
algebra with product

\begin{equation}
( A, B )   \;=\;   p \times A \times B  +  (1 - p ) \times B \times A ,
\label{eq:one-1}
\end{equation}
where $p$ is a parameter, $A$, $B$ are matrices or operators (hereon
assumed
to be Hermitean), and $A\times B$ is the conventional associative
product.

It is easy to see that the above product is indeed jointly Lie-
and Jordan-admissible because $[A, B]U \;=\; (1 - 2p)\times (A\times B -
B\times A)$  and $\{A, B\}_U \;=\; A\times B + B\times A$. However,
there
exist no (finite) value of p under which product (\ref{eq:one-1})
recovers the Lie
product. As a result, product (\ref{eq:one-1}) cannot be used for
possible
coverings of current physical theories.

In view of the above occurrence, Santilli  introduced in 1967 [2] a new
notion
of Lie-admissibility which is Albert's definition [loc. cit.], plus the
condition that the algebras $U$ admit Lie algebras in their
classification or,
equivalently, that the generalized Lie product admits the conventional
one as
a particular case.

        As an illustration, we introduced the the algebra with
product (Ref.  [2], Eq.(8), p. 573)

\begin{equation}
    ( A, B )  \;=\;  p \times A \times B  -   q \times B \times A ,
\label{eq:one-2}
\end{equation}
and related time evolution in the infinitesimal and finite forms $(\hbar
\;=\;=
 1)$ [3,4]

\[
  i \times dA / dt  \;=\;  p \times A \times H  q \times H \times A ,
\]
\begin{equation}
  A(t)  \;=\; \{ e^{i \times q \times H \times t}\} \times A(0) \times
\{
 e^{-i \times p \times t \times H} \} ,
\label{eq:one-3}
\end{equation}
where: $p$ and $q$ are non-null parameters with non-null values $p\pm
q$. It is
easy to see that product (\ref{eq:one-2}) is Lie- and Jordan-admissible
and
admits the Lie and Jordan products as particular (nondegenerate) cases.

        Structures (\ref{eq:one-2}) and (\ref{eq:one-3}) resulted to be
insufficient for physical applications because, as we shall see in Sect.
3,
the parameters $p$ and $q$ become operators under the time evolution of
the
        theory. We therefore introduced in 1978 [5] (see also monograph
[6] of
1983) the notion of {\it general Lie-admissibility} which is the notion
of
ref. [1] plus the conditions that algebras $U$ admit {\it Lie-isotopic} 
[5,6]
(rather then Lie) algebras in their attached antisymmetric form and
admit
ordinary Lie algebras in their classification.

The latter notion was realized via the $general Lie-admissible
product$ (first introduced in ref. [5b], p. 719; see Ref. [6] for a more
detailed treatment)

\begin{equation}
( A, B )  \;=\;   A \times P \times B  -  B \times Q \times A ,
\label{eq:one-4}
\end{equation}
and time evolution in infinitesimal and finite forms (Ref. [5b], pp.
741, 742, and [6])

\[
 i \times dA / dt  \;=\;  A \times P \times H  -  H \times Q \times A ,
\]
\begin{equation}
 A(t)  \;=\; \{ e^{i \times H \times Q \times t}\} \times A(0) \times
\{ e^{-i \times t \times P \times H}\} , 
\label{eq:one-5}
\end{equation}
where $H$ is Hermitean but $P$ and $Q$ are nonsingular, generally
nonhermitean
matrices or operators with non-singular values $P\pm Q$ admitting of the
parametric values p and q as particular cases. The conventional
Heisenberg's equations are evidently recovered for $P \;=\; Q \;=\; 1$.

Note that the attached products $[A, B]_U \;=\; (A, B) - (B, A) \;=\;
A\times T\times B - B\times T\times A,$ $T \;=\; P + Q$, and $\{A, B\}_U
\;=\;
(A, B) + (B, A) \;=\; A\times T\times B + B\times T\times A$, $T \;=\; P
- Q$,
are still Lie and (commutative) Jordan, respectively, although of a more
general type called {\it isotopic} [5,6].
Note also that the $P$ and $Q$ operators must be sandwiched in between
the
elements $A$ and $B$ to characterize an algebra as commonly understood
in
mathematics [5,6].  It should be finally indicated that, when properly
written, Hamilton's equations with external terms possess precisely a
Lie-admissible structure.

        In 1989 Biedenharn [7] and Macfarlane [8]  introduced the
so-called
{\it q-deformations}, with a structure of the type

        \begin{equation}
        A \times B  -  B \times A     \;\rightarrow\;     A \times B
- q \times A \times B ;   
\label{eq:one-6}
\end{equation}
which were followed by a number of papers so large to discourage an
outline (see, e.g., representative papers [9]). More recently, other
types of deformations of relativistic quantum formulations appeared in
the literature under the name of {\it k-deformations} (see, e.g., Ref.s
[10],
{\it quantum groups}  (see, e.g. Ref.s [11]) and others generalizations.
It is
evident that theories of type(\ref{eq:one-6}) are a particular case of
the
broader Lie-admissible theories (\ref{eq:one-2}) and (\ref{eq:one-4}).

        Unfortunately, even thouygh mathematically impeccable, all the
above theories have resulted to possess a number of physiocal
shortcomings
investigated in Ref.s [19,23]. As a necessary condition to exit the
class of equivalence of quantum mechanics, Lie-admissible theories,
q-deformations, k-deformations, quantum groups, and all that must have a
{\it nonunitary time evolution}, $U\times U^\dagger \;\not=\; I$. When
the
these theories are formulated on conventional spaces over conventional
fields,
the following physical shortcomings are simply unavoidable: 
\begin{itemize}

\item[(1)] {\it Lack of invariance of the fundamental unit} (that of the
enveloping operator algebra), because under nonunitary transforms we
have $I
\;\rightarrow\; I\prime \;=\; U\times I\times U^\dagger \;=\; U \times
U^\dagger
\;\not=\; I$. This implies lack of invariance of the basi= c units of
space
and time, with consequential lack of unambiguous application of the
theories
to experiments, because it is not possible to conduct a meaningful
measurement, say, of a length, with a stationary meter changing  in
time.

\item[(2)] {\it Lack of conservation of the Hermiticity in time}, with
consequential lack of physically acceptable observables (see Sect. 3 for
more
details).

\item[(3)] {\it Lack of invariance of physical laws}, e.g., because of
the
lack of invariance of the deformed brackets under the time time
evolution of
the theory.

\item[(4)] {\it Lack of uniqueness and invariance of numerical
predictions},
        because of the lack of uniqueness (e.g., in the exponentiation)
and
invariance (e.g., of special functions and transforms) needed for data
elaboration (for instance, the ''q-parameter'' becomes a ''Q-operator''
under
a nonunitary transform, $Q \;=\; q\times U\times U^\dagger$, with
consequential evident loss of all original  special functions and
transforms
constructed for the q-parameter).

\item[(5)] {\it Loss of the axioms of the special relativity}, an
occurrence
of all generalizations under consideration, evidently because deformed
spaces
        and symmetries are no longer isomorphic to the original ones.
This
creates the sizable problems of: first, identifying new axioms capable
of
replacing Einstein's axioms; second, proving their axiomatic
consistency; and,
third, establishing them experimentally.  \end{itemize}

In this note we shall present a conceivable resolution of the
        above physical shortcomings based on the use of a new
mathematics
called {\it genomathematics}, as recently identified in memoir [12]. To
render
the note self-sufficient, we shall first outline in Sect. 2 the
rudiments of
the genomathematics and then indicate in Sect. 3 the invariant
formulation of
$(p, q)$- and $(P, Q)$-deformations.

The reader should keep in mind that the most  serious shortcoming of
generalized theories under consideration in this note is the loss of
Einstein's axioms. Our primary objective is therefore to attempt the
formulation of generalized theories in such a way to {\it preserve} the
axioms
of the special relativity, although in generalized spaces and fields. 
If
achieved, this result will be sufficient, alone, to resolve all possible
physical shortcomings.

\section{GENOMATHEMATICS}
\setcounter{equation}{0}
The main idea of the Lie-admissible theory~[6] is that its structure
(\ref{eq:one-5}) is inherent in the {\it conventional} Lie theory. In
fact, a
one-parameter connected Lie group realized via Hermitean operators $X
\;=\;
X^\dagger$ on a Hilbert space $\cal H$ has in reality the structure of a
{\it bi-module} (also called in nonassociative algebras {\it spit-null }
extension,see, e.g., Ref. [13]).

In nontechnical terms, the structure of a Lie group as a
bimodule is essentially characterized by an action from the left $U^>$
and an
action from the right $^<U$ with explicit realization and
interconnecting
conjugation

\begin{eqnarray}
A(t)  &\;=\;&  U^> \times Q(0)  \times ^<U  \;=\;  \{ e^{i\times
X^{>\times
w}} \} > A(0) < \{ e^{-i\times w\times^<X} \} \;=\;\nonumber \\
&\;=\;&  ( I^> +  i \times X^> \times w + ... )  > A(0)  < (^<I -
i\times w\times^<X + ... ) ,\nonumber \\
U^>  &\;=\;&  ( ^<U )^\dagger  \;=\; U^>,\;\;\; X^> \;=\; (^<X
)^\dagger  \;=\;
X, \;\;\; \hat{I}^>  \;=\;  ^<I  \;=\; I ,
\label{eq:two-1}
\end{eqnarray}
(where $w$ is a Lie parameter and the multiplications $>$ and $<$
represent
conventional associative products ordered to the right and to the left,
respectively). The infinitesimal version in the neighborhood of the unit
then acquires the familiar form

\begin{equation}
i [ A(dw)  -  A(0) ) / dw  \;=\;  A < X  -  X > A   \;=\;
A \times X  -  X \times A,        
\label{eq:two-2}
\end{equation}
which clarifies that in the product $AxX \;=\; A<X (X\times A \;=\;
X>A)$, $X$
in actuality acts from the right (from the left).

The bimodular structure is generally ignored in the conventional
formulation
of Lie's theory because un-necessary. In fact, in a Lie bimodule
$\{^<\cal H,
\cal H^>\}$, where $^<\cal H \;=\; \cal H^> \;=\; \cal H$ is a
conventional
Hilbert space, the modular action to the right and to the left are
interconnected with the simple bimodular rules~[14]  $X^\times > \psi^>
\;=\;
X\times\psi \;=\; -^<\psi<^<X \;=\; -\psi\times X$, where $\psi^> \in
\cal
H^>$, $^<\psi \in ^<\cal H$, $X^>$ is an element of the universal
enveloping
associative algebra $\xi^>(L)$ of the considered Lie algebra $L \approx
[\xi^>(L)]^-$ for the action to the right and $^<X \in ^<\xi(L)$~[14].
Since
$\cal H^> \;=\; ^<\cal H \;=\; \cal H$, and $\xi^>(L) \;=\; ^<\xi(L)
\;=\;
\xi(L)$.  The {\it bireprsentations} of the bimodular structure
$\{^<\xi(L),
\xi^> (L)\}$ over $\{^<\cal H, \cal H^>\}$ can then be effectively
reduced to
the {\it one-sided representations}, or just {\it representations} for
short,
of $\xi(L)$ over $\cal H$, as well known. However, as we shall see
shortly,
the original bimodular structure of Lie's theory is no longer trivial
for
broader realizations of axioms (\ref{eq:two-1}).

Lie-admissible structure (\ref{eq:one-5}) was proposed  [5b] on the
basis of
the mere observation that the abstract axioms of the bimodular structure
(\ref{eq:two-1}) do not necessarily require that the multiplications $>$
and
$<$ must be conventional, because they can also be generalized, provided
that
they remain {\it associative}.  In other word, the abstract axiomatic
structure of the action from the right, $U^>>A(0)$ is that of a {\it
right
modular associative action}, with no restriction on the realization of
the
associative product, and the same occurs for the action from the left
$A(0)<^<U$.

The simplest possible broadening of the Lie  version is given by the
{\it isotopies
of Lie's theory}, first proposed in Ref.s [5], then studied in various
works
(see Ref. [6] for a comprehensive presentation as of 1983), and it is
called
the {\it Lie-Santilli isotheory} (see, e.g., Refs.  [15-18]). It is
essentially
characterized by the lifting of the conventional right modular
associative
product $U^>>A(0) \;=\; U^> \times T \times A(a)$ with conjugate from
the left
$A(0)<^<U$, where $T \;=\; T^\dagger$ is a fixed, well behaved, nowhere
singular and Hermitean matrix or operator of the same dimension of the
considered representation . Its inverse $\hat{I} \;=\; T^{-1}$ is then a
fully
acceptable, generalized, left and right unit, $I \times A \;=\; A >
\hat{I}
\;=\; \hat{I} < A \;=\; A < \hat{I} \;=\; \hat{I}$ for all possible
elements
$A$.

The isotopies then require, for mathematical and physical consistency,
the reconstruction of the {\it entire} Lie theory with respect to the
new unit
$\hat{I}$ and isoproduct $> \;=\; < \;=\; \hat{\times}$, including:
numbers and fields; vector, metric and Hilbert spaces; Lie algebras,
groups
and symmetries; transformation and representation theories; etc.
[15-18]. This
intermediate level of study also possesses a trivial bimodular
structure, in
the sense that its two-sided representations can be effectively reduced
to the
one-sided form.

Following the prior achievement of sufficient mathematical
maturity in Ref.  [12], the physical profiles of the isotopic
realization of
axioms (\ref{eq:two-1})  have been studied in details in the recent
memoir [19],
including most importantly the resolution of problematic aspects (1)-(5)
of
the preceding section. A knowledge of Ref. [19] is useful for a
technical
understanding of this note.

Our objective is the realization of the abstract  axioms of bimodular
structure (\ref{eq:two-1}) via the generalized associative laws
originally submitted in
        Ref. [5b] of 1978, under the name of {\it genoassociative
multiplication
and unit (or genomultiplication and genounit} for short), then studies
in Ref.s
[6,20], and more recently studied in details in Ref. [12],

\[
A > B  \;=\;  A \times P \times B ,\;\;\; A < B \;=\;  A \times Q \times
B ,
\]
\[
\hat{I}^>  \;=\;  P^{-1} ,\;\;I^> > A  \;=\;  A > \hat{I}^>  \;=\;  A ,
\;\;
^<I= \;=\;  Q^{-1},\;  ^<I < A  \;=\;  A < ^<I  \;=\;  A ,
\]
\begin{equation}
\hat{I}^>  \;=\;  P^{-1}  \;=\;  ( ^<\hat{I} )^\dagger  \;=\; Q^\dagger,
\label{eq:two-3}
 \end{equation}
where $P \;\not = \; Q$ are well behaved, everywhere invertible,
nonhermitean
matrices or operators generally realized via real-valued nonsymmetric
matrices of the same dimension of the considered Lie representation.
Moreover, it is requested that that $P+Q$ and $P-Q$ are nonsingular to
preserve a well defined Lie and Jordan content, respectively. To
differentiate forms (\ref{eq:two-3}) from the isotopic ones, I called
them
{\it genotopic} in Ref. [5], to denote their character of inducing a
more
general realization. $I^>$ and $^<I$ and then called {\it genotopic
units} and
$P$ and $Q$ the {\it genotopic elements}.

Broader products and units (\ref{eq:two-3}) characterize  the following
more general
realization of the abstract axioms (\ref{eq:two-1}) I tentatively called
{\it
Lie-admissible transformation group} [5,6,12,20]

\begin{eqnarray}
A(t)  \;&=&\;  U^> > A(0)  < ^<A  \;=\;  \{ e_>^{i\times X\times w}\} >
A(0) < \{ <e^{-i\times w \times X} \}  \;=\; \nonumber \\
\;&=&\;  \{ e^{i\times X\times P\times w}\} \times A(0) \times
\{e^{-i\times w\times Q\times X} \} \;=\; \nonumber \\
\;&=&\; ( I +  i \times X \times P \times w + . . . )  \times A(0) <
(^<I -
i\times w\times ^<Q \times ^<X + . . . ) , \nonumber \\
\;&=&\;  ( I^> +  i \times X \times w + . . . ) > A(0)  < (^<I-i  \times
w
\times ^<X  + . . . ) , \nonumber
\end{eqnarray}
\[
U^> \;=\; (^<U )^\dagger ,\;\; X^> \;=\; (^<X )^\dagger \;=\; ^<X  \;=\;
X,\;\; P^> \;=\;  P  \;=\;  (^<Q )^\dagger  \;=\;  Q^\dagger ,
\]
\begin{equation}
\hat{I}^>  \;=\;   P^{-1} \;=\;  (^<I)^\dagger \;=\;  ( Q^\dagger)^{-1},
\label{eq:two-4}
\end{equation}
with infinitesimal version in the neighborhood of the genounits
characterized by the {\it general Lie-admissible algebra} [loc. cit.]

\begin{equation}
i \times [ A(dw)  -  A(0) ) / dw  \;=\;  A < X  -  X > A   \;=\;
A \times P \times  X  -  X  \times Q \times A,
\label{eq:two-5}
\end{equation}
 where we have used the {\it genoexponentiation} to the {\it right} and
{\it to the
left} [12,18]

\[
e_>^{i\times X\times w}  \;=\;  I>  +  i\times X\times w / 1! +
(i\times X\times w) > (i\times X\times w) / 2 ! + . . .  \;=\;
\{e^{i\times X\times P\times w} \} \times I^>,
\]
\begin{equation}
_<e^{i\times w\times X} \;=\;  ^<I  +  i\times X\times w / 1 !  +
(i\times X\times w) > (i\times X\times w) / 2 ! + . .  .  \;=\;
^<I \times \{ e^{i\times w\times Q\times X} \} ,
\label{eq:two-6}
\end{equation}

It is at this point where the essential bimodular character of
axioms (\ref{eq:two-1}) acquire their full light because no longer
effectively reducible
to a one-sided form. It is evident that realization (\ref{eq:two-4}) and
(\ref{eq:two-5}) of the conventional Lie axioms (\ref{eq:two-1})
coincides
with the Lie-admissible equations (\ref{eq:one-5}) and (\ref{eq:one-4}).
For
this reason, realizations (\ref{eq:two-3})-(\ref{eq:two-6}) are assumed
as the
foundation of the Lie-admissible theory under study in this section.

 The central assumption we are studying herein is the bimodular
lifting of the unit of Lie's theory $I \;\rightarrow\; \{^<,\hat{I},
\hat{I}^>\}, ^<\hat{I} \;=\; (\hat{I}^>)^\dagger$. To achieve
consistency,
the {\it entirety} of the Lie theory must be lifted into a dual
genotopic
form, with no known {\it exception}. A rudimentary review of the
emerging
{\it genotopic mathematics} or {\it genomathematics} for short of Ref.
[12]
plus unpublished aspects is the following.

\vskip 0.5cm
{\sl DEFINITION 1 [21]:} Let $F \;=\; F(a,+,\times)$ be a conventional
field
of (reaL $R$, complex $C$ or quaternionic $Q$) numbers   a   with
additive
unit 0, multiplicative unit $I \;=\; 1$, sum $a+b$ and product $a\times
b$.
The {\it genofields to the right} $F^> \;=\; F^>(a^>,+^>,\times^>$) are
rings
with elements $a^> \;=\; a\times I^>$ called {\it genonumbers}, where  
$a$
is an element of $F$, $\times$ is the multiplication in $F$, and $I^>
\;=\;
P^{-1}$ is a well behaved, everywhere invertible and non-Hermitean
quantity
generally outside $F$, equipped with all operations {\it ordered to the
right,
i.e., the ordered genosum to the right, ordered genoproduct to the
right,}
etc.,

\begin{equation}
(a^>) +^>(b^>)  \;=\;  (a+b)\times I^>,\;\; (a^>)\times^>(b^>) \;=\;
(a^>)>(b^>) (a^>)\times P\times (b^>) \;=\; (a\times b)\times I^>,
\label{eq:two-7}
\end{equation}
{\it genoadditive unit to the right} $0^> \;=\; 0$ and {\it genounit to
the
right} $I^>$.  The {\it genofields to the left} $^<F \;=\;
^<F(^<a,^<+,^<\times)$ are rings with genonumbers $^<a \;=\; ^<I\times
a$, all
operations ordered to the left, such as genosum $(^<a)^<+(^<b) \;=\;
^<b)
\;=\; ^<I\times (a+b)$, genoproduct $(^<a)<(^<b) \;=\; (^<a)\times
Q\times
(^<b) \;=\; ^<I\times (a\times b)$, etc., with {\it additive genounit to
the
left } $^<0 \;=\; 0$ and {\it multiplicative genounit to the left} $^<I
\;=\;
Q^{-1}$ which is generally different than than the genounit $I^>$ to the
right. A {\it bigenofield} is the structure $\{^<F, F^>\}$ with
corresponding
bielements, biunits, bioperations, etc. holding jointly to the left and
right
under the condition $\hat{I}^> \;=\; (^<I)^\dagger$.

\vskip 0.5cm
{\sl LEMMA 1 [21]:} Each individual genofield to the right $F^>$ or to
the left $^<F$ is a field isomorphic to the original field $F$. Thus,
the
liftings $F \;\rightarrow \; F^>$, $F \;\rightarrow\; ^<F$ and $\{F, F\}
\;\rightarrow\; \{^<F, F^>\}$ are axiom-preserving.

\vskip 0.5cm
{\sl REMARKS:} In the definition of fields (and isofields [21]) there
is no ordering of the multiplication in the sense that in the products
$a\times b$ and $a\hat{\times}b \;=\; a\times T\times b$, $T \;=\;
T^\dagger$,
one can either select a multiplying $b$ from the left, $a<b$ or $b$
multiplying $a$ from the right $a>b$, because $a>b \;=\; a<b$ (even for
non-commutative isofields such as the isoquaternions). A genofield
requires
that all multiplications and related operations (division, moduli, etc.)
be
ordered {\it either} to the right {\it or} to the left because now, for
a
commutative field $F \;=\; R$ or $C$, we have the properties $a>b \;=\;
b>a$
and $a<b \;=\; b<a$, but in general $a>b \;=\; a\times P\times b \;\not
=\;
a<b \;=\; a\times Q\times b$.  Note that in each case the {\it genounit}
is
the {\it left and} right unit, Eq.s (\ref{eq:two-3}).  The important
advances
of Ref.  [21] are therefore the identification, first, that the axioms
of a
field remain valid when the multiplication is ordered to the right or to
the
left, and, second, each ordered multiplication  can be generalized,
provided
that it remains associative. The above mathematical occurrences permit
the
axiomatization of irreversibility beginning with the most fundamental
quantities, units and numbers. In fact, the unit and product to the
right,
$I^>$ and $>$, characterize {\it motion forward in time} while the
conjugate
quantities $^<I$ and $<$ characterize {\it motion backward in time}.
Irreversibility is then ensured under the condition $I^> \;\not =\; ^<I$
because all subsequent mathematical structures, being always built on
numbers,
must preserve the same axiomatization of irreversibility, as a necessary
condition for consistency.

\vskip 0.5cm
{\sl DEFINITION 2 [12]:} Let $S \;=\; S(r,g,R)$ be a conventional
n-dimensional metric or pseudo-metric space with local chart $r \;=\;
\{r^k\}$, $k \;=\; 1, 2, ..., n$, nowhere singular, real-valued and
symmetric
metric $g \;=\; g(r, ...)$ and invariant $r^2 \;=\; r^t\times g\times r$
(where $t$ denotes transposed) over a conventional real field $R \;=\;
R(a,+,\times)$.  The n-dimensional {\it genospaces to the right} $S^>
\;=\;
S^>(r^>,G^>,R^>)$ are vector spaces with local {\it genocoordinates to
the
right } $r^> \;=\; r\times I^>$, {\it genometric} $G^> \;=\; P\times
g\times
I^> \;=\; (g^>)\times I^>, g^> \;=\; P\times g$, and {\it genoinvariant
to the
right}

\begin{equation}
( r^>)^{2>}  \;=\;  ( r^> )^t > ( G^> ) > r^>  \;=\;  [ r^t \times ( g^>
)
\times r ] \times I^> \in R^> ,
\label{eq:two-8}
\end{equation}
which, for consistency, must be a genoscalar to the right with structure
$n\times I^>$ and be an element of the genofield $R^>$ with common
genounit
to the right $I^> \;=\; P^{-1}$ where $P$ is given by an everywhere
invertible, real-valued, non-symmetric nxn matrix. The n-dimensional
{\it genospaces to the left} $^<S \;=\; ^<S(^<r,^<+,^<F)$ are genospaces
over
genofields with all operations ordered to the left and a common
nxn-dimensional genounit to the left $^<I \;=\; Q^{-1}$ which is
generally
different than that to the right but verifying the interconnecting
condition
$P \;=\; Q^\dagger$. The {\it bigenospaces} are the structures $\{^<S,
S^>\}$
with bigenocoordiantes, etc, defined over the bigenofield $\{^<R, R^>\}$
under
the condition $I^> \;=\; (^<I)^\dagger$.

\vskip 0.5cm
{\sl LEMMA 2 [12]:} Genospaces to the right $S^>$ and, independently,
those to the left $^<S$ (thus bigenospaces $\{^<S, S^>\}$) are locally
isomorphic to the original spaces $S$ ( $\{S, S\}$ ).

PROOF. The original metric $g$ is lifted in the form $g \;\rightarrow\;
P\times g$, but the unit is lifted by the {\it inverse} amount $I
\;\rightarrow\; I^> \;=\; P^{-1}$ thus preserving the original axioms
(because
the invariant is $({\rm length})^2\times ({\rm unit})^2)$, and the same
occurs
for the other cases.  q.e.d.

\vskip 0.5cm
{\sl REMARKS.} The best way to see the local iso\-mor\-phism be\-tween
con\-ven\-tio\-nal   and ge\-no\-spa\-ces
is by nothing that the latter are the results of
the following novel degree of freedom of the former (here expressed for
the
case of a scalar complex function $P$)

\begin{eqnarray}
r^t \times g \times r \times I &\equiv& r^t \times g \times r \times Q
\times
Q^{-1} \equiv   ( r^t \times g^> \times r ) \times I^>  \equiv \nonumber
\\
&\equiv&  P^{-1} \times P \times ( r^t \times g \times r \times I) 
\equiv
^<I \times ( r \times ^<g \times r^t )
\label{eq:two-9}
\end{eqnarray}
which is another illustration of the structure of the basic invariant of
metric spaces $[{\rm length}]^2\times [{\rm Unit}]^2$.

\vskip 0.5cm
{\sl DEFINITION 3 [12]:} The {\it genodifferential calculus to the
right} on
a genospace $S>(r^>, R^>)$ over $R^>$ is the image of the conventional
differential calculus characterized by the expressions (where we have
ignored for notational simplicity the multiplication to the right by
$I^>$)

\[
dr^k \;\rightarrow\; d^>r^k \;=\;  (I^>)^k_i \times dr^i ,
dr_k \;\rightarrow\; d^>r_k \;=\;  P_k^i \times dr_i ,
\]
\begin{equation}
\partial / \partial r^k  \;\rightarrow\;  \partial^> / \partial^>r^k 
\;=\;
P_k^i \times \partial / \partial r^i ,\;\; \partial / \partial r_k
\;\rightarrow\; \partial^> / \partial^>r_k  \;=\;  I^k_i \times  I /
\partial r_i ,
\label{eq:two-10}
\end{equation}
with all operations ordered to the right and main properties

\begin{equation}
\partial^>  r^i / \partial^> r^j  \;=\;  \delta i_j  ,\;\; \partial^>
r_i /
\partial^> r_j   \;=\;  \delta_i^j ,\;\;{\rm  etc.}
\label{eq:two-11}
\end{equation}
The {\it genodifferential calculus to the left} is the conjugate of the
preceding one for the genounit to the left $^<I \;\not =\; I^>$. The
{\it bigenodifferential calculus} is that acting on $\{^<S, S^>\}$ over
$\{^<R, R^> \}$ for $I^> \;=\; (^<I)^\dagger$.

\vskip 0.5cm
{\sl LEMMA 3 [12]:} The genocalculus to the right and, independently,
that to the left preserve all original properties, such as commutativity
of the
second-order derivative, etc.

\vskip 0.5cm
{\sl REMARKS.} A important advance of Ref. [12] is the identification
of an insidious lack of invariance where one would expect it the least,
in the
conventional differential calculus, because traditionally formulated
without indicating its dependence on the selected unit. As a result, all
{\it generalized} equations of motion expressed in terms of {\it
conventional}
derivative, e.g., $dA/dt$, {\it are not} invariant.

\vskip 0.5cm
{\sl DEFINITION 4 [12]:} The {\it genogeometries to the right}, or {\it
to the
left}, or the {\it bigenogeometries} are the geometries of the
corresponding
genospaces when entirely expressed via the applicable geonomathematics,
including the genodifferential calculus.

\vskip 0.5cm
{\sl LEMMA 4 [loc. cit.]:} The {\it ge\-no\-eucli\-de\-an,
ge\-no\-min\-kow\-ski\-an,
ge\-no\-rie\-man\-ni\-an} and {\it ge\-no\-symp\-lec\-tic
geo\-met\-ri\-es  to the right} and,
independently, to the left and their combined bimodular form, are
locally
isomorphic to the original geometries (i.e., the verify their abstract
axioms).

\vskip 0.5cm
REMARKS. Another intriguing property identified in memoir [12]
is that {\it the Riemannian axioms do not necessarily need symmetric
metrics}
because the metrics can also be {\it nonsymmetric} with structure $g^>
\;=\;
P\times g$, $P \;=\; P^t$ real-valued but nonsymmetric, provided that
the
geometry is formulated on a genofield with genounit given by the {\it
inverse
} of  the nonsymmetric part, $I^> \;=\; P^{-1}$, and the same occurs for
the
case to the left.  This property has permitted the first quantitative
studies
on the {\it irreversibility} of interior gravitational problems via the
conventional {\it Riemannian axioms} [20], e.g., the geometrization of
the
irreversible black hole model by Ellis, Nonopoulos and Mavromatos
[21],which
has precisely a Lie-admissible structure, and other models.  These
remarks are
important to begin to see the physical relevance of Biedenharn's
q-deformations when written in an axiomatically correct form.

\vskip 0.5cm
{\sl DEFINITION 5 [12]:} Let $\cal H$ be a conventional Hilbert space
with
states $|\psi >$, $| \varphi >$, $...$, inner product $< \varphi |
\times |
\psi >$ over the field $C \;=\; C(c,+,\times)$ of complex numbers and
normalization $< \psi | \times | \psi > \;=\; 1$. A {\it genohilbert
space to the
right} $\cal H^>$ is a right genolinear space with genostates $|
\psi^>>$, $|
\varphi^>>$, $...$, {\it genoinner product and genonormalization to the
right}

\begin{equation}
< \varphi^> | > | \psi^> >  \;=\;  < \psi^> | \times P \times | \psi^>
> \times I^> \;\in\; C^>(c^>+^>,\times^>),\;\;  < \varphi^> | > | \psi^> >
\;=\; I^>
\label{eq:two-12}
\end{equation}
defined over a genocomplex field to the right $C^>(c^>,+^>,\times^>)$
with a
common genounit $I^> \;=\; P^{-1}$. A {\it genohilbert space to the
left}
$^<\cal H$ is the left conjugate of $\cal H^>$ with left genounit $^<I
\;=\;
Q^{-1}$ generally different than $I^>$.  A {\it bigenohilbert space} is
the
bistructure $[^<\cal H, \cal H^>\}$ over the bigenofield $\{^<C, C^>\}$
under
the conjugation $I^> \;=\; (^<I)^\dagger$.

\vskip 0.5cm
{\sl LEMMA 5:} The right-, left- and bi-genohilbert spaces are locally
isomorphic to the original space $\cal H$.

{\sl PROOF.} The original inner product is lifted by the amount $<
|\times|
> \;\rightarrow\; < |\times P\times | >$, but the underlying unit is lifted by
the {\it inverse} amount, $1 \;\rightarrow\; P^{-1}$, thus leaving the
original
axiomatic structure unchanged. q.e.d.

\vskip 0.5cm
{\sl REMARK.} The understanding of genooperator theory requires the
knowledge   that it is a consequence of the following, hitherto unknown
degree
of freedom of conventional Hilbert spaces (where $P$ is independent from
the
integration variable for simplicity)

\begin{eqnarray}
< \varphi | \times | \psi >  &\equiv&  < \psi  | \times | \psi > \times
P
\times P^{-1}  \equiv  < \varphi | \times P \times | \psi > \times
P^{-1}
\;=\; < \varphi | < |  \psi > \times ^<I  \equiv \nonumber \\
&\equiv&   < \varphi | \times | \psi > \times Q \times Q^{-1}  \equiv <
\varphi  | > | \psi > \times I^>,
\label{eq:two-13}
\end{eqnarray}
which sis evidently the Hilbert space counterpart of the novel
invariance (\ref{eq:two-9}). It should be noted that new invariances
(\ref{eq:two-9}) and (\ref{eq:two-13}) have remained undetected since
Riemannian's and Hilbert's times, respectively, because they required
the
prior discovery of {\it new numbers}, those with an arbitrary, generally
nonhermitean unit.

\vskip 0.5cm
{\sl DEFINITION 6:}  {\it Genolinear operators to the right} are
operators
$A$, $B$, $...$, of a genoenveloping algebra to the right verifying the
condition of genolinearity (i.e., linearity on $\cal H^>$ over $C^>$),
and a
similar occurrence holds for the left case. In particular, we have the
{\it genounitary operators to the right and to the left}

\begin{equation}
U^> > U^{>\dagger}  \;=\;  U^{>\dagger} > U  \;=\;I^>,\;\; ^<U <
^<U^\dagger
\;=\; ^<U^\dagger < ^<U \;=\;  ^<I .
\label{eq:two-14}
\end{equation}
   When applied on the bistructure $\{^<\cal H, \cal H^>\}$ over $\{^<C,
C^>\}$, the theory is {\it bigenolinear.}

\vskip 0.5cm
{\sl LEMMA 6:} Operators X which are originally Hermitean on $\cal H$
over $C$
remains Hermitean on $\cal H^>$ over $C^>$, or on $^<\cal H$ over $^<C$
(i.e.,
genotopies preserve the original observables).

PROOF. The condition of genohermiticity on $\cal H^>$ reads $X^{\dagger
>}
\;=\; Q\times Q^{-1}\times X\dagger\times Q\times Q^{-1} \;=\;
X\dagger$.
q.e.d.

\vskip 0.5cm
{\sl LEMMA 7:} Under sufficient topological conditions, any
conventionally nonunitary operator on $\cal H$ can be identically
written in a
genounitary form to the right or to the left.

PROOF. Any operator $U$ of the  considered class such that $U\times
U^\dagger
\;\not =\; I$ can always be written

\begin{equation}
U  \;=\;  ( U^> ) \times Q^{1/2}  \;\;  {\rm or}\;\;    P^{1/2}\times
(^<U ) ,
\label{eq:two-15}
\end{equation}
and properties (\ref{eq:two-14}) follows. q.e.d.

\vskip 0.5cm
{\sl REMARKS.} The reader should be aware that the entire theory of
linear operators on a Hilbert spaces must be lifted into a genotopic
form for
consistency. For instance, conventional operations, such as TrX, DetX,
etc. can be easily proved to be inapplicable for genomathematics, and
must be replaced with the corresponding genoforms. The same happens for
{\it all} conventional and special functions and transforms. A
systematic
study of the theory of genolinear operators will be conducted elsewhere.

        We are now equipped to present, apparently for the first time,
the central notion of this note which consists of the old notion of
Lie-admissibility upgraded with the systematic use of genomathematics.

\vskip 0.5cm
{\sl DEFINITION 7:} Consider the conventional Lie theory with ordered
N-dimensional basis of Hermitean operators $X \;=\; \{X_k\}$, parameters
$w
\;=\; \{w_k\}$, universal enveloping associative algebra $\xi \;=\;
\xi(L)$,
Lie algebra $L \approx [\xi(L)]^-$, corresponding, (connected) Lie
transformation group $G$ on a space $S(r,F)$ with local coordinates $r
\;=\;
\{r^k\}$ over a field $F$.

The {\it Lie-admissible theory} (also called  {\it Lie-Santilli
genotheory}
[15-18]) is here defined as a step-by-step bimodular lifting of the
conventional Lie theory defined on bigenospaces over bigenofields, and
includes:

(5.A) The  {\it universal genoenveloping as\-so\-cia\-ti\-ve al\-ge\-bra
to the right}
$\xi^>(L)$ of an N-di\-men\-sio\-nal Lie algebra $L$ with ordered basis
$X^> \equiv
X \;=\; \{X_k\},\: k \;=\; 1, 2, ..., N$,  genounit $I^> \;=\; Q^{-1}$,
genoassociative product $X_i>X_j \;=\; X_i\times Q\times X_j$ and
infinite-dimensional genobasis characterized by the {\it genotopic
Poincare'-Birkhoff-Witt theorem to the right}

\begin{equation}
I^> \;=\; Q^{-1},\;\; X_k,\;\; X_i > X_j \;( i \leq j ),\;\; X_i> X_j >
X_k
(i \leq j \leq k ), ...
\label{eq:two-16}
\end{equation}
and genoexponentiation (\ref{eq:two-16}); the {\it universal
genoassociative
algebra to the left} $^<\xi(L)$ with genounit $^<I \;=\; P^{-1}$ and
genoproduct $X_i<X_j \;=\; X_i\times P\times X_j$, with
infinite-dimensional
genobasis characterized by the {\it genotopic Poincare'-Birkhoff-Witt
theorem
to the left}

\begin{equation}
^<I \;=\; P^{-1},\;\; X_k,\;\; X_i < X_j \;( i \leq j ),\;\; X_i< X_j <
X_k
\;(i \leq j \leq k ), ...
\label{eq:two-17}
\end{equation}
and genoexponentiation to the left (\ref{eq:two-6}); the {\it
bigenoenvelope}
is the bistructure $\{^<\xi, \xi^>\}$ defined on corresponding
bigenospaces
and bigenofields under the condition $I^> \;=\; (^<I)^\dagger$.

(5a) A {\it Lie-Santilli genoalgebra} is a bigenolinear bigenoalgebra
defined on $\{^<\xi, \xi^>\}$ over $\{^<F, F^>\}$ with Lie-admissible
product

\begin{equation}
( X_i , X_j )  \;=\;  X_i < X_j  -  X_j  > X_i  \;=\;  X_i \times P
\times
X_j  - X_j  \times Q \times X_i .
\label{eq:two-18}
\end{equation}

(5c)  A (connected) {\it Lie-Santilli genotransformation group} is the
biset $\{^<G, G^>\}$ of bigenotransforms on $\{^<S, S^>\}$ over $\{^<F,
F^>\}$
with genounits $^<I \;=\; (I^>)^\dagger$

\begin{eqnarray}
r^{>}\prime\;&=&\;  ( U^> ) > r^>  \;=\;  ( U^> ) \times Q > r \times
I^>  \;=\;
 V \times r \times I^> ,\;\; U^> \;=\; V \times I^>, \nonumber \\
^<r \prime\;&=&\;  ^<r < ( ^<U )  \;=\;  ^<I \times r \times P \times (
^< U )
\;=\; ^<I \times r \times  W , \;\;^<U    ^<I \times W ,
\label{eq:two-19}
\end{eqnarray}
verifying the following conditions: genodifferentiability of the maps
$G^>>S^> \;\rightarrow\; S^>$ and $^<S \;\leftarrow\; ^S < ^<G$,
invariance of
the genounits and genolinearity, with realizations $U^> \;=\;
exp_>(i\times w\times X)$ and $^<U \;=\; exp_<(-i\times w\times X)$,
genolaws

\begin{equation}
U^>(w^{.>} > U^>(w^> {}\prime) \;=\; U^>(w^> + w^> {}\prime),\;\;  
U^>(w^>) >
U^>(-w^>)  \;=\;  U^>(0^>)  \;=\;  I^>.
\label{eq:two-20}
\end{equation}
and Lie-admissible algebra  in the neighborhood of the
genounits $\{^<I, I^>\}$ according to rule (\ref{eq:two-4}).

\vskip 0.5cm
{\sl LEMMA 8:} Lie-admissible product (\ref{eq:two-18}) verifies the
{\it Lie}
axi\-oms when de\-fi\-ned on  $\{^<\xi, \xi^>\}$ over $\{^<F, F^>\}$.

{\sl PROOF.} The genoenvelopes to the left $^<\xi$ and to the right
$\xi^>$ are
isomorphic to the original envelope $\xi$, thus implying $^<I(A<B) \;=\;
(AS>B)_{I>}$ i.e., the value of the genoproduct $A<B \;=\; A\times
P\times B$,
when measured with respect to the genounit $^<I \;=\; P^{-1}$, is equal
to
that of the genoproduct $A>B \;=\; A\times Q\times B$ measured with
respect to
the genounit $I^> \;=\; Q^{-1}$. q.e.d.

        The most important property of this section, which is an evident
consequence of the preceding analysis, can be expressed as follows:

\vskip 0.5cm
{\sl THEOREM:} Lie-admissible groups as per Definition 7 coincide at
the abstract level with the original Lie-transformation groups.

\vskip 0.5cm
{\sl REMARKS.} Note that the generators of the original Lie
algebra are not lifted under genotopies, evidently because they
represent
conventional physical quantities, such as energy, linear momentum,
angular
momentum, etc. Only the {\it operations} defined on them are lifted.
Note also
that, when conjugation $P \;=\; Q^\dagger$ is violated, the Lie axioms
are
lost.  Note also that the genotheory is highly nonlinear, because the
elements
$P$ and $Q$ in genotransforms (\ref{eq:two-19}) have un unrestricted
functional dependence, this including that in the local coordinates.
Nevertheless, genomathematics reconstructs linearity in genospaces over
genofields. The same happens for nonlocality, noncanonicity,
nonunitarity and
irreversibility [20].  In fact, on genospaces over genofields,
genotheories
are fully linear, local, canonical unitary and reversible. Departures
from
these axiomatic properties occur only in their {\it projection} over
conventional; spaces and fields. These are evident fundamental
conditions to
lift nonlinear, nonlocal, noncanonical, nonunitary and irreversible
theories
into a form compatible with the notoriously linear, local, canonical,
unitary
and reversible axioms of the special relativity.

        Needless to say, we have been able to present in this note only
the rudiments of the needed genomathematics, with the understanding that
its
detailed study is rather vast indeed. Also, by no means, genomathematics
should be considered as the most general possible form admitted by the
Lie axioms. Mathematics and physics are disciplines which will never
admit ''final theories''. In fact, a still broader multivalued
hyperrealization of Lie's theory has already been identified in Ref.
[12] and cannot be treated here for brevity.

\section{INVARIANT FORMULATION OF q - DE\-FOR\-MA\-TI\-ONS}
\setcounter{equation}{0}
We are now equipped to submit the suggested invariant formulation of the
(p, q)- [2] or q-deformations [2,7,8]. First, we have to identify the
following insufficiencies:

\begin{itemize}
\item[(I)] No invariant formulation is possible for (p, q)-parameters
because, under the nonunitary time evolution of the theory, brackets
(\ref{eq:one-2}) or (\ref{eq:one-8}) assume the general Lie-admissible
form
(\ref{eq:one-4}) (for which reason the latter was submitted in the first
place
[5b,6]),

\[
U \times (A, B) \times U^\dagger  \;=\; p \times U \times A \times B
\times
U^\dagger -  q \times U \times B \times A \times U^\dagger  \;=\;
A\prime  \times P
\times B\prime   -  B\prime \times Q \times A\prime  ,
\]
\begin{equation}
P  \;=\;  p \times ( U \times U^\dagger )^{-1},\;  Q  \;=\;  q \times (
U
\times u^\dagger)^{-1},\;  A\prime  \;=\;  U \times A \times
U^\dagger.\;  B\prime   \;=\;
U \times B \times U^\dagger .
\label{eq:three-1}
\end{equation}

\item[(II)] Despite such a generality, the formulation are still not
physically acceptable because they generally violate the crucial
conjugation  $P  \;=\;  Q^\dagger$, without which there is the loss of
the Lie
axioms (Sect. 2) with consequential problems in invariance, causality,
etc.
The condition $P \;=\; Q^\dagger$ is therefore assumed hereon.

\item[(III)] Brackets  $(A, B) \;=\; A\times P\times B - B\times Q\times
A, P
\;=\; Q^\dagger$ on conventional spaces and fields are still not
invariant
and, therefore, they have all problematic aspects (1)-(5) of the (p, q)-
and
q-deformations  (Sect.  1). In fact, under an additional (necessarily)
nonunitary transform we have
\end{itemize}

\[
U \times (A, B) \times U^\dagger  \;=\;  U \times A \times P \times B
\times
U^\dagger -  U \times B \times A \times U^\dagger  \;=\;  A\prime 
\times P\prime
\times B\prime   -  B\prime \times Q\prime  \times A\prime  ,
\]
\begin{equation}
P\prime  \;=\; U^{\dagger-1}\times P \times U^{-1},\;\;     Q\prime 
\;=\;
U^{\dagger-1}\times Q\times U^{-1},\;\;    A\prime  \;=\;  U \times A
\times
U^\dagger \;\; B\prime  \;=\;  U \times B \times U^\dagger .
\label{eq:three-2}
\end{equation}
This implies the lack of invariance of the fundamental genounits $I^>
\;=\;
P^{-1}$ and $^<I \;=\; Q^{-1}$, with consequential ambiguous physical
applications.

The only possible resolution of the above problematic aspects
known to this author is the formulation of the q-parameter deformations
in the
operator (P, Q)-deformations formulated via the genomathematics of Sect.
2, i.e., on bigenofield, bigenospaces, bigenoalgebra, etc.

 In fact, it is easy to see that each structure to the right is
invariant under the action of the genogroup to the right, e.g.,
$U^>>I^>>U^{>\dagger} \;=\; I^>$, $U^>(A>B)>U^{>\dagger} \;=\; A\prime
>B$, the
initial genohermiticity to the right can be proved to remain invariant
under
the action of a genogroup to the right, etc.

>From these grounds, genominkowskian spaces, the genopoincare' symmetry and the genospecial
relativity are expected to {\it coincide} at abstract level with the
conventional corresponding structures, with the understanding that the
detailed study of this expectation will be predictably long and cannot
possibly be done in this note.

We close with a simple rule for the explicit construction of
invariant (P, Q)-deformations and related genomathematics. It is based
on the
systematic use of two nonunitary transforms for the characterization of
motion forward and backward in time,

\begin{equation}
A \times A^\dagger  \;\not =\; I,\;\;   B \times B^\dagger  \;\not =\; 
I,
\;\; A \times B^\dagger  \;=\;  I^> \;=\;  Q^{-1},\;\;   B\times
A^\dagger
\;=\; ^<I \;=\; ( I^>)^\dagger .
\label{eq:three-3}
\end{equation}

It is then easy to see that the {\it entire} genomathematics of the
preceding section follows via a simple application of the above two
transforms.
For instance, the genonumbers to the right are given by the above
transforms of conventional numbers  $A\times > a\times B^\dagger \;=\;
a\times (A\times B^\dagger) \;=\; A\times I^>$, the genoproduct to the
right
is given by the same transform  $A\times (A\times B)\times B^\dagger
\;=\;
A\prime \times Q\times B{}\prime $, $Q \;=\;  (A\times B^\dagger)^{-1}$
with the correct
Hermiticity properties, etc.

Most importantly, the Lie-Santilli genogroups  and genoalgebras can also
be
derived via the above dual nonunitary map. In fact, a conventional,
right
modular Lie group is lifted under the transform $A\times B^\dagger$ into
the
forward genoform

\begin{eqnarray}
\lefteqn{e^{i \times X \times w}   \;\rightarrow\;     A \times \{ e^{i
\times
X \times w} \} \times B^\dagger \;=\;} \nonumber \\
&\;=\;&  A \times ( I + i \times X \times w / 1 !  +  ( i \times X
\times w)
\times ( i \times X \times w) / 2 ! + ... ) \times B^\dagger
\;=\;\nonumber \\
&\;=\;&  I^>  +^>  i^>>X^>>w^>/^>1 !^>  + \nonumber \\
&& \;\;+ ^>  ( i^>>X^>>w^> ) > ( i^>>X^> >w^>
) > ( i^>>X^> >w^> /^> 2!^> +^> . . . )  \;=\; \nonumber \\
&\;=\;&  ( I  +  i \times X^> \times Q \times w ) / 1 !  +  \nonumber \\
&&\;\;+ ( i \times X^>
\times Q \times w ) \times ( i \times X^> \times Q \times w ) / 2 ! + .
. . )
\times I^>  \;=\; \nonumber \\
&\;=\;&  \{ e^{i \times X^> \times w} \} \times I^>  \equiv  e_> i
\times
 X \times w, \nonumber \\
I^>& \;=\;& A\times B^\dagger \;=\; Q^{-1},\;\; \nonumber \\
X^>&\;=\;&  A \times X \times B^\dagger,\;\;
w^> \;=\; w \times I^>,\;\;  i^>>X^> >w^> \equiv i \times X^> \times w ,
\label{eq:three-4}
\end{eqnarray}

with a conjugate lifting for the left modular action. Lie-admissible
algebras then follows in the neighborhood of the genounits $\{^<I,
I^>\}$.

In conclusion, Biedenharn's 1989 paper on q-deformations [7],
when expressed in an invariant, (P, Q)-operator, Lie-admissible form,
deals
with one of the most important problems of the physics of this century,
the {\it origin of irreversibility}. In fact, the invariant formulation
of the
deformations permits the identification of the origin of irreversibility
at the ultimate level of physical reality, such as a proton in the core
of a collapsing star [22]. In this case all conventional,
action-at-a-distance, potential forces are represented via the
conventional Hamiltonian H, while contact, zero-range, nonhamiltonian,
irreversible effects are represented via the forward isounit $I^> \;=\;
A\times B^\dagger$ with different backward from $^<I \;=\; (^<I) \;=\;
B\times A^\dagger$. When applicable in interior problems, the emerging
theory
is then {\it structurally irreversible}, that is, irreversible even for
reversible Hamiltonians.

\vskip 1cm
\centerline{\bf \large ACKNOWLEDGMENTS}
I would like to express my gratitude to H. Tributsch, C. A. C. Dreismann
and L. Pohlmann for the opportunity of visiting the Hahn-Meitner
Institute in Berlin in September 1995 and for long, detailed and
critical examinations of the various aspects of Lie-admissible theories.



\newpage
\begin{thebibliography}{00}
\bibitem[1.]     A. A. Albert, {\it Trans. Amer. Math. Soc.}  64, 552
(1948)

\bibitem[2.]      R. M. Santilli, {\it Nuovo Cimento} 51, 570 (1967)

\bibitem[3.]      R. M. Santilli, {\it Suppl. Nuovo Cimento} 6, 1225
(1968)

\bibitem[4.] R. M. Santilli, {\it Meccanica} 1, 3 (1969) [6b];

\bibitem[5.]      R.M. Santilli, {\it Hadronic J.} 1, 224 [5a], 574 [5b]
and1267 [5c] (1978); {\it Phys. Rev.} D 20, 555 (1979) [5d]

\bibitem[6.] R. M. Santilli, {\it foundations of Theoretical Mechanics},
Vol.
II:  {\it Birkhoffian Generalization of   Hamiltonian Mechanics}
(Springer-Verlag, Heidelberg-New York, 1983)

\bibitem[7.]      L. C. Biedernharn {\it J. Phys.} A 22, L873 (1989)

\bibitem[8.]      A. J. Macfarlane, {\it J. Phys.} A 22, L4581 (1989)

\bibitem[9.]      V. Dobrev, in {\it Proceedings of the Second Wigner
Symposium}, Clausthal 1991 (World     Scientific, Singapore, 1992). J.
Lukierski, A. Novicki, H.  Ruegg and V. Tolstoy, Phys. Lett. B     264,
331
(1991). O. Ogivetski, W.B.  Schmidke, J. Wess and B. Zumino, {\it Comm.
Math.
Phys. } 50,  495 (1992). S.  Giller, J. Kunz, P. Kosinky, M. MajewskiI
and P.
Maslanka, {\it Phys. Lett.  B.} 286, 57      (1992)

\bibitem[10.]     J. Lukierski, A. Nowiski and H. Ruegg, {\it Phys.
Lett. B}
293, 344 (1992). J. Lukierski, H. Ruegg  and W. R\"{u}hl, {\it Phys.
Lett. B }
313, 357 (1993). J. Lukierski and H. Ruegg, {\it Phys. Lett. B} 329, 189
(1994).  S.  Majid and H. Ruegg, {\it Phys. Lett. B} 334, 348 (1994)

\bibitem[11.]     T. L. Curtis, B. Fairlie and Z.K. Zachos, Editors,
{\it Quantum
Groups} (World, Scientific,     Singapore, 1991). Mo-Lin Ge and Bao Heng
Zhao,
Editors,  {\it Introduction to Quantum Groups        and Integrable
Massive
Models of Quantum Field Theory} (World Scientific , Singapore, 1991).  
Yu. F.
Smirnov and R. M. Asherova, Editors, {\it Proceedings of the Fifth
Workshop
Symmetry Methods in Physics} (JINR, Dubna, Russia, 1992)

\bibitem[12.]     R. M. Santilli, {\it Rendiconti Circolo Matematico
Palermo,
Suppl.} 42, 7 (1996)

\bibitem[13.]     R. D. Schafer, {\it An Introduction to Nonassociative
Algebras} (Academic Press, New York, 1966)

\bibitem[14.]     R. M. Santilli, {\it Initiation of the representation
theory of
Lie admissible algebras on bimodular        Hilbert spaces, Hadronic J.}
3, 440
(1979)

\bibitem[15.]     D. S. Sourlas and G. T. Tsagas, {\it Mathematical
Foundations of
the Lie-Santilli Theory} (Ukraine    Academy of Sciences, Kiev, 1993)

\bibitem[16. ]    J. L$\hat{o}$hmus, E. Paal and L. Sorgsepp, {\it
Nonassociative
Algebras in Physics} (Hadronic Press, Palm   Harbor, FL, 1994) [5d]

\bibitem[17.]     J. V. Kadeisvili, {\it Santilli's Isotopies of
contemporary
Algebras, Geometries and Relativities}     (Hadronic Press, FL, 1991,
Second
Edition,
Ukraine Academy of Sciences, Kiev , in press).

\bibitem[18.]     J. V. Kadeisvili, {\it Math. Methods in Applied
Sciences},
19, 362 (1996)

\bibitem[19.]     R. M. Santilli, {\it Relativistic hadronic mechanics:
Nonunitary
axiom-preserving completion of  relativistic quantum mechanics}, in
press
(1997)

\bibitem[20.]     R. M. Santilli, {\it Elements of Hadronic Mechanics},
Vol.s
I and II, (Ukraine Academy of Sciences,   Kiev, Second Ediityibn, l995)

\bibitem[21.]     R. M. Santilli, {\it Algebras, Groups and Geometries}
10, 273
(1993)

\bibitem[ 22.]    J. Ellis, N. E. Mavromatos and D. V. Nanopoulos in
{\it Proceedings of the Erice Summer        School, 31st Course: From
Superstrings
to the Origin of Space-Time}, World Scientific (1996)

\bibitem[23]    D. F. Lopez, in {\it Symmetry Methods in Physics} 
(Memorial
Volume dedicated to Ya. S.     Smorodinsky), A. N. Sissakian, G. S.
Pogosyan
and
S. I. Vinitsky, Editors, J.I.N.R., Dubna,       Russia (1994), p. 300;
and
Hadronic J. 16, 429 (1993); Jannussis and D. Skaltzas, {\it Ann. Fond.
L. de
Broglie} 18, 1137 (1993); A. Jannussis, R. Mignani and R. M. Santilli,
{\it Ann. Fond. L. de Broglie} 18,    371 (1993);  D. Schuch, {\it Phys.
Rev.
A}, 55 (1997), in press; R. M. Santilli, '' Problematic aspects       
of
classical
and quantum deformations'', Preprint IBR-TH-97-S-037, submitted for
publication

\end{thebibliography}
\end{document}